\documentclass[twocolumn,showpacs,superscriptaddress,preprintnumbers,amsmath,amssymb,prb]{revtex4}

\usepackage{graphicx}
\usepackage{epsfig}

\begin{document}

\title{Role of oxygen in the electron-doped superconducting cuprates}

\author{J. S. Higgins}
\affiliation{Center for Superconductivity Research, Department of
Physics, University of Maryland, College Park, Maryland, USA
20742-4111}
\author{Y. Dagan}
\affiliation{Center for Superconductivity Research, Department of
Physics, University of Maryland, College Park, Maryland, USA
20742-4111}
\affiliation{School of Physics and Astronomy, Raymond
and Beverly Sackler Faculty of Exact Sciences, Tel Aviv University,
Tel Aviv 69978, Israel}
\author{M. C. Barr}
\affiliation{Center for Superconductivity Research, Department of
Physics, University of Maryland, College Park, Maryland, USA
20742-4111}
\author{B. D. Weaver}
\affiliation{Naval Research Laboratory, Code 6818, Washington, DC,
USA 20375}
\author{R. L. Greene}
\affiliation{Center for Superconductivity Research, Department of
Physics, University of Maryland, College Park, Maryland, USA
20742-4111}

\date{\today}

\begin{abstract}
We report on resistivity and Hall measurements in thin films of the
electron-doped superconducting cuprate
Pr$_{2-x}$Ce$_{x}$CuO$_{4\pm\delta}$. Comparisons between x = 0.17
samples subjected to either ion-irradiation or oxygenation
demonstrate that changing the oxygen content has two separable
effects: 1) a doping effect similar to that of cerium, and 2) a
disorder effect. These results are consistent with prior
speculations that apical oxygen removal is necessary to achieve
superconductivity in this compound.
\end{abstract}

\maketitle

A striking property of the high-temperature cuprate superconductors
is that the superconducting transition temperature (T$_{c}$) depends
on the number of carriers put into the copper oxygen planes (i.e.
doping). In the electron-doped (\textit{n}-doped) cuprate system
RE$_{2-x}$Ce$_{x}$CuO$_{4}$ (RE = La, Pr, Nd, Sm, Eu), Ce$^{4+}$
partially replaces the rare earth ion (RE$^{3+}$) thereby
introducing electrons into the CuO$_{2}$ plane.~\cite{Takagi}
However, for these materials doping alone is insufficient: oxygen
reduction is a necessary step to achieve superconductivity. Usually,
this reduction process is done by annealing the sample in a low
pressure oxygen environment. Oxygen has a strong effect not only on
T$_{c}$, but also on many other properties such as the resistivity
and Hall effect,~\cite{Jiang} and the temperature at which
antiferromagnetic order is established.~\cite{Kang} Understanding
why oxygen reduction is vital for superconductivity, and why it has
such a strong effect on the transport and other normal state
properties of \textit{n}-doped cuprates, is the focus of this paper.

Of the ideas put forth to explain the role of oxygen reduction in
the \textit{n}-doped cuprates, the predominant explanations are: to
decrease impurity scattering,~\cite{Xu} to suppress the long-range
antiferromagnetic order in the CuO$_{2}$ planes,~\cite{Richard,
Riou} or to change the number of mobile charge
carriers.~\cite{Jiang2} Oxygen in RE$_{2}$CuO$_{4}$ occupies two
sites in the ideal lattice: sites in the CuO$_{2}$ plane and in the
PrO layer. In practice, a small percent of oxygen ($\approx$ 1 $\%$)
is also found to occupy a third, impurity site (the apical site)
located directly above the copper in the CuO$_{2}$ plane. Therefore,
a decrease in impurity scattering and the appearance of
superconductivity is consistent with the view that oxygen is removed
from the apical sites. Indeed, neutron scattering experiments found
an average apical oxygen reduction of $\simeq$ 0.06 per formula
unit~\cite{Radaelli} in undoped Nd$_{2}$CuO$_{4}$. By contrast,
Richard \textit{et al}.~\cite{Richard} and Riou \textit{et
al.}~\cite{Riou} reported that, for superconducting samples, oxygen
reduction occurs primarily in the CuO$_{2}$ planes and they
suggested that this suppresses long-range antiferromagnetic ordering
in the plane and allows the competing phase of SC to appear. In the
context of charge carriers, Jiang \textit{et al}.~\cite{Jiang2}
found oxygen reduction to increase the number of hole-like carriers
in a two-band model of these materials.

\begin{figure}
\centerline{\epsfig{file=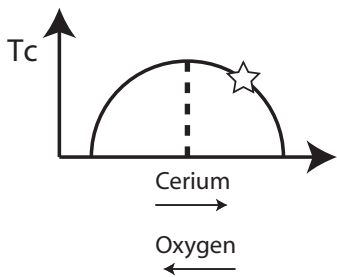,width=1.75in}}
\caption{T$_{c}$ versus doping phase diagram (schematic) for
Pr$_{2-x}$Ce$_{x}$CuO$_{4-\delta}$. Dashed line indicates optimal
cerium doping (x $\simeq$ 0.15). The star on the diagram indicates
the Ce doping presented in this paper. Electron carrier
concentration increases to the right. Increasing oxygen
concentration decreases the electron carrier concentration and is
represented by moving to the left. Upon cerium doping, the material
becomes more metallic and the residual resistivity \emph{decreases}
as one moves to the right in the diagram.} \label{PhaseDiagram}
\end{figure}

When a metal is cooled down to low temperatures, the electrical
resistivity is dominated by impurity (or disorder) scattering. The
residual resistivity, $\rho_{0}$, is inversely proportional to the
carrier density and the time between impurity scattering events. The
Hall coefficient, R$_{H}$, at low temperatures is primarily related
to the number of carriers. It is therefore possible to differentiate
disorder effects from carrier concentration effects by measuring
both $\rho_{0}$ and R$_{H}$ at low temperatures. Our experiment
looks at one of the \textit{n}-doped cuprates,
Pr$_{2-x}$Ce$_{x}$CuO$_{4\pm\delta}$ (PCCO), and is based on a
comparison between oxygenated thin films and optimally prepared thin
films subjected to ion-irradiation, primarily a source of disorder.
One should bear in mind that if oxygen has a doping effect, then
adding oxygen should add holes into the CuO$_{2}$ planes, and thus
act in a manner opposite to cerium doping. Most prior transport
studies as a function of oxygenation~\cite{Jiang,Jiang2} and
irradiation~\cite{Weaver,Woods,Woods2} were performed on samples
having optimum Ce doping, x $\simeq$ 0.15. In that case, adding
disorder (by irradiation) or holes (by adding oxygen) would have the
same effect on T$_{c}$ and $\rho_{0}$, i.e. T$_{c}$ would decrease
and $\rho_{0}$ would increase as the material became either
disordered or ``underdoped'', and the results would be nearly
indistinguishable from each other (see Fig.~\ref{PhaseDiagram}). It
is therefore more informative to study the overdoped region where
adding holes would result in an \emph{increase} in T$_{c}$ and an
increase in $\rho_{0}$ as illustrated in Fig.~\ref{PhaseDiagram}.
The results of our analysis in this paper show that changing the
oxygen concentration in overdoped PCCO can be described by two
additive effects: one due to a change in carrier concentration, and
another due to disorder.

Thin films of c-axis oriented, overdoped PCCO (x = 0.17) were
deposited from a stoichiometric target onto (001) oriented
SrTiO$_{3}$ substrates using a pulsed laser deposition technique. An
LPX 300, 248 nm KrF excimer laser provided a fluence of 1.5 - 2
J/cm$^{2}$ at a frequency of 10 Hz, yielding $\simeq$ 0.3 $\AA$ per
pulse. The substrate temperature was maintained at $\approx$ 770
$^{\circ}$C in a 230 mTorr N$_{2}$O environment, inside a vacuum
chamber. Annealing (oxygen reduction) was performed post-deposition,
in situ, at $\approx$ 720 $^{\circ}$C in a low pressure N$_{2}$O
environment. The time of the annealing was adjusted to give a sharp
and symmetric transition in the imaginary component of
ac-susceptibility measurements, which we used as a measure of sample
quality. The typical full width at half maximum was better than 0.2
K for all the thin films in this study. The N$_{2}$O pressure was
varied depending on the desired result. The film subjected to
ion-irradiation was prepared using ``optimal annealing'' conditions,
which started at a pressure of 1x10$^{-4}$ Torr and decreased down
to 3x10$^{-5}$ Torr, where the pressure was maintained for the
remainder of the annealing. In order to increase the oxygen content
for the oxygenated films, we increased the N$_{2}$O pressure
relative to the ``optimal annealing'' pressure and maintained this
pressure for the entire annealing process. This increase in total
pressure results in an increase in the oxygen partial pressure which
corresponds to an increase in the oxygen population relative to
optimally reduced films. The time of the anneal is then adjusted in
order to minimize the width of the superconducting transition in
ac-susceptibility measurements. All annealing times for either
method were between 10 and 17 minutes. The oxygenated films were
annealed in 1 x 10$^{-4}$, 1 x 10$^{-3}$, and 2.3 x 10$^{-1}$ Torr
of N$_{2}$O. The film thicknesses were $\approx$ 3000 $\AA$, as
determined by Rutherford Backscattering Spectroscopy. All films were
patterned into a Hall bar geometry using either photolithography or
a mechanical mask and ion-milling. The oxygenated films had the
typical eight contact pad Hall geometry. The film used in
irradiation was patterned such that up to six Hall bars could be
irradiated separately along a shared current path. This sample was
irradiated with 2 MeV H$^{+}$ ions in doses of 0, 1, 2.5, 8, and 32
x 10$^{15}$ ions/cm$^{2}$. Transport measurements were performed in
a Quantum Design Physical Property Measurement System in fields up
to 14 T and in temperatures down to 350 mK. Cerium overdoped samples
(x = 0.17) were chosen to maximize the chance of observing an
increase in T$_{c}$ caused by increasing the oxygen content, as
mentioned above.

\begin{figure}
\centerline{\epsfig{file=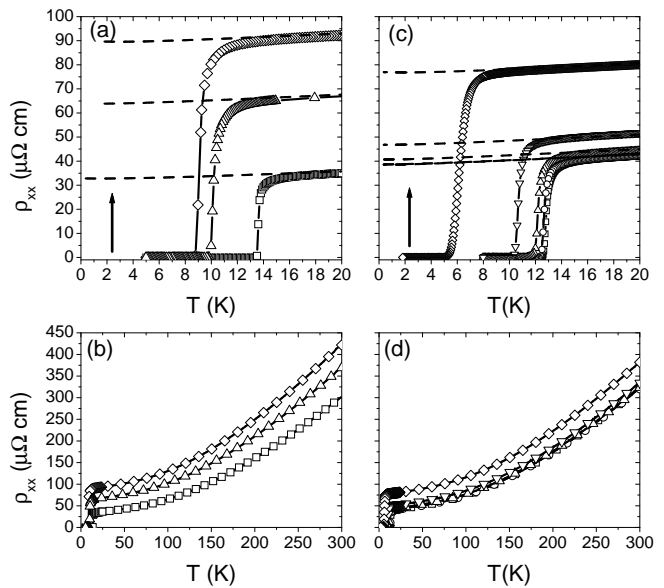,clip=,silent=,width=3.375in}}
\caption{ab-plane resistivity versus temperature for x = 0.17
cerium-doped PCCO films. Symbols are data taken in zero applied
magnetic field. Dashed lines are taken in 10 T field (H $\parallel$
c-axis) and coincide with H = 0 T above T$_{c}$. Scales for (a) and
(c) are the same, as well as for (b) and (d). (a) Films with
different oxygen content. The arrow indicates the order of
increasing oxygen, and the corresponding post-deposition annealing
pressures are: 1 x 10$^{-4}$ ($\square$), 1 x 10$^{-3}$
($\bigtriangleup$), and 2.3 x 10$^{-1}$ Torr ($\lozenge$). (b) Full
temperature scale for the same oxygenated films presented in (a).
(c) A single, optimally annealed, film subjected to increasing
irradiation doses. The arrow indicates the order of increasing
irradiation corresponding to doses 0 ($\square$), 1 ($\bigcirc$),
2.5 ($\bigtriangleup$), 8 ($\bigtriangledown$), and 32 ($\lozenge$)
x 10$^{15}$ ions/cm$^{2}$. (d) Full temperature scale for the
irradiated film presented in (c).} \label{ResFig}
\end{figure}

Fig.~\ref{ResFig} shows the ab-plane resistivity measurements as a
function of temperature for both the oxygenated and irradiated
films. Measurements were performed in zero magnetic field and also
in 10 T field, applied parallel to the c-axis. We determined the
transition temperature from the peak in the derivative plot
($\frac{d\rho}{dT}$ vs T) of the zero field data. The residual
resistivity was calculated from the normal state data measured in 10
T, using the relation $\rho=\rho_{0}+AT^{\beta}$ with $\rho_{0}$, A,
and $\beta$ as the free parameters. Here we were interested in
extracting $\rho_{0}$. The oxygenated samples (Fig.~\ref{ResFig}(a))
show an increase in $\rho_{0}$ and a decrease in T$_{c}$ with
increasing oxygen content. The irradiated sample
(Fig.~\ref{ResFig}(c)) shows the same behavior with increasing dose,
however the change in T$_{c}$ for a given change in $\rho_{0}$ is
larger than in the oxygenated samples.

\begin{figure}
\centerline{\epsfig{file=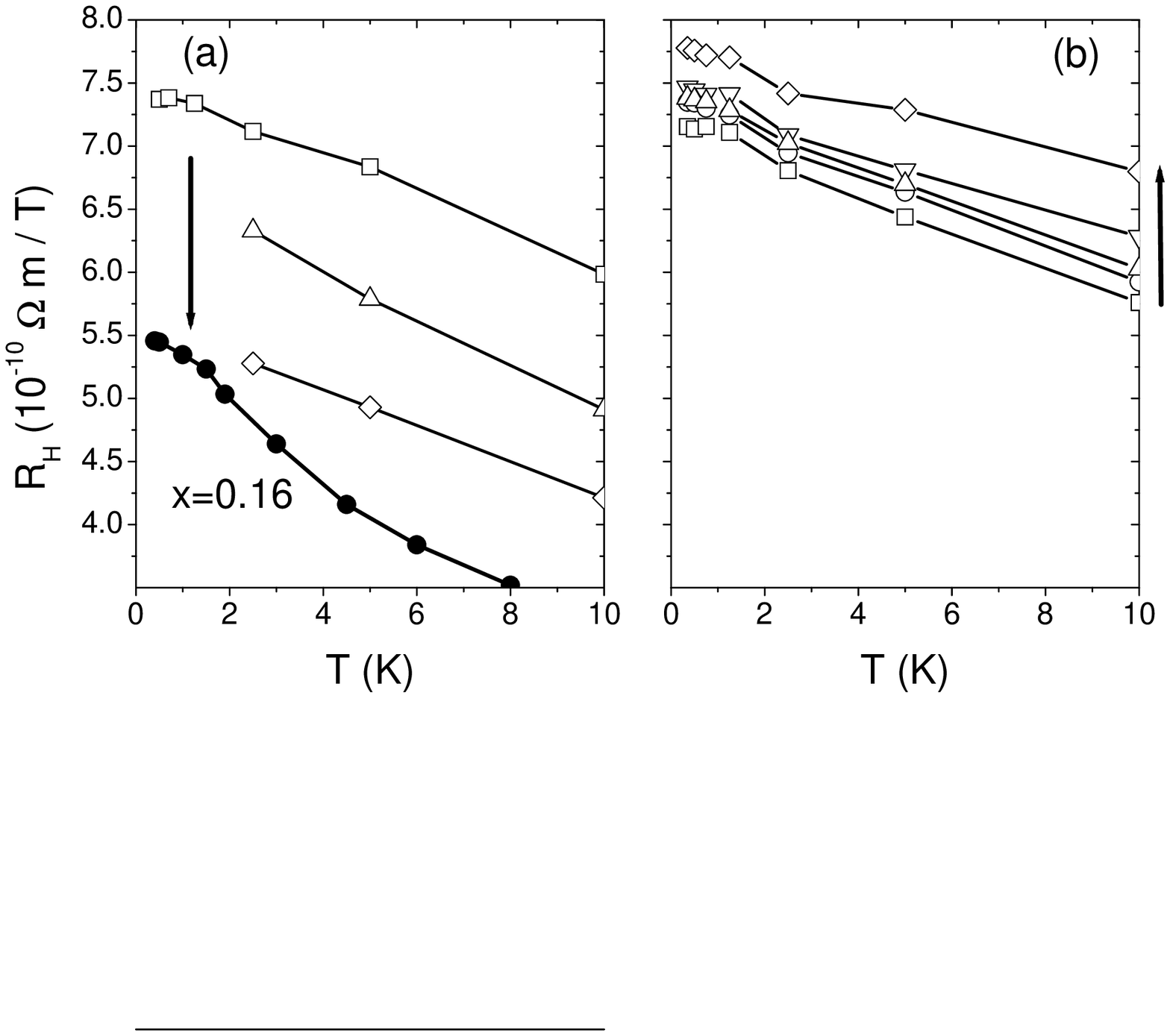,clip=,silent=,width=3.375in}}
\caption{Hall coefficient (R$_{H}$) versus temperature for the same
films in Fig.~\ref{ResFig}. Scales for both plots are identical.
Although this material is classified as n-doped, R$_{H}$ is
\emph{positive} on the overdoped side of the doping phase
diagram~\cite{Dagan}. (a) Films with different oxygen content (arrow
indicates order of increasing oxygen).  The data labeled x = 0.16
($\bullet$) is from an optimally annealed x = 0.16 cerium-doped thin
film. (b) Single (optimally annealed) film subjected to increasing
irradiation doses (arrow indicates order of increasing irradiation
dose).} \label{HallFig}
\end{figure}

In Fig.~\ref{HallFig}, we show the results of the Hall measurements.
For the temperature range shown, field sweeps of $\pm$14 T were used
to determine the Hall coefficient (R$_{H}$). The oxygenated samples
show a decrease in magnitude of R$_{H}$ as the oxygen content
increases. The trend is consistent with a decrease in cerium doping,
from x = 0.17 toward x = 0.16. T$_{c}$, however, does not increase
as would be expected from a purely cerium-doping standpoint
(Fig.~\ref{PhaseDiagram}). In contrast, the irradiated sample shows
an \emph{increase} in R$_{H}$ with irradiation. Since the relative
change in R$_{H}$ of the irradiated sample is in the direction
opposite to that observed in the oxygenated samples, we make the
assumption that the primary result of irradiation is to induce
disorder (i.e. affect scattering) with no effect on the carrier
density. This allows us to use the irradiation data as a measure of
only the disorder for these samples. This assumption draws on the
fact that 2 MeV H$^{+}$ ion-irradiation mainly creates oxygen
vacancies and interstitials, with no loss of total oxygen. This
rearrangement of the oxygen should have a minor effect on the
carrier density.~\cite{Woods,Tolpygo}

The above mentioned observations lead us to the model chosen to
analyze the data and to ultimately clarify the role of oxygen in
this class of material. The residual resistivity is given by
$\rho_{0}$ = $\frac{m^{*}}{ne^{2}\tau}$ where m$^{*}$ is the
effective mass, n is the carrier density, e is the electronic
charge, and $\tau$ is the time between elastic scattering events.
The Hall coefficient for a simple metal is given as R$_{H}$ =
$\frac{1}{ne}$. PCCO is usually not classified as a simple metal and
its transport properties have been qualitatively interpreted in
terms of a two-band model.~\cite{Fournier,Gollnik} However, using a
two-band model, without expanding the number of measurements, makes
quantitative analysis dubious. Thus we restrict ourselves to the
one-band Drude model and bear in mind that this model is
oversimplified. While we do not calculate the carrier density from
R$_{H}$, we do use R$_{H}$ as an empirical measure of carrier
concentration. And, since R$_{H}$ is related to the number of
carriers and $\rho_{0}$ to both the number of carriers and impurity
scattering, it is possible to differentiate disorder effects from
the carrier concentration effects by measuring $\rho_{0}$ and
R$_{H}$ at low temperatures.

Using these assumptions, we now determine the contribution to
$\rho_{0}$ and T$_{c}$ due to additional disorder in the oxygenated
samples. We then compare the disorder effect on T$_{c}$ with the
measured T$_{c}$ and we show that the oxygenated samples have an
additional, positive contribution to T$_{c}$, which has a behavior
similar to that of cerium-doping. In order to show this, we write
the residual resistivity of the oxygenated samples as:
\begin{equation}\label{residual}
  \rho_{0}(O_{2})=\frac{m^{*}}{e^{2}(n+\Delta n)}(\frac{1}{\tau_{0}}+\frac{1}{\tau_{1}})
\end{equation}
where $\Delta$n represents any change in the carrier density,
$\tau_{0}$ represents the low temperature elastic scattering term
inherent in the optimally annealed system, and $\tau_{1}$ represents
the low temperature elastic scattering term due to additional
disorder introduced by the extra oxygen. The effective mass is taken
to be independent of doping~\cite{Padilla,Zimmers} and is a constant
in this analysis. After expanding the ($\frac{1}{n+\Delta n}$)
factor, we rewrite equation~\ref{residual} as
$\Delta\rho_{0}(O_{2})$ by subtracting $\rho_{0}$.
\begin{equation}\label{deltaresidual}
\Delta\rho_{0}(O_{2})=\frac{m^{*}}{ne^{2}}[-\frac{\Delta
n}{n\tau_{0}}+\frac{1}{\tau_{1}}(1-\frac{\Delta n}{n})]
\end{equation}
The first term represents changes in the oxygenated samples due only
to changes in the carrier concentration. The second term contains
effects from both additional disorder and carrier concentration. To
simplify equation~\ref{deltaresidual}, we rewrite it as:
\begin{equation}\label{deltaresidual2}
  \Delta\rho_{0}(O_{2})=\Delta\rho_{0}(R_{H})+\Delta\rho_{0}(disorder)
\end{equation}
where we use R$_{H}$ as a measure of carrier concentration.

\begin{figure}
\centerline{\epsfig{file=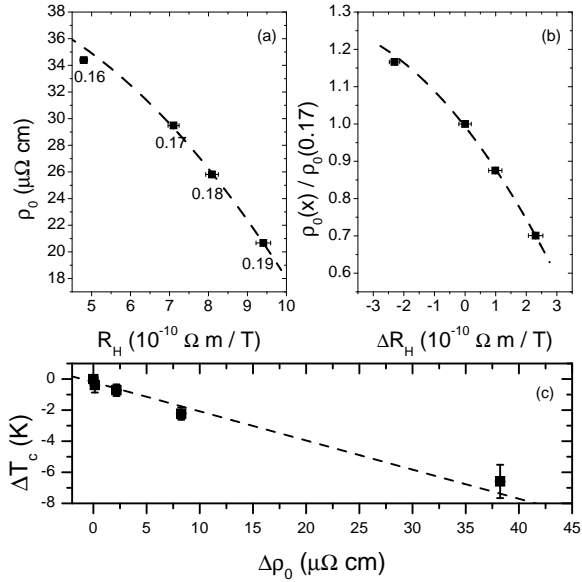,clip=,silent=,width=3in}}
\caption{(a) Plot of $\rho_{0}$ versus R$_{H}$ at T = 2.5 K for
optimally annealed cerium-doped samples. Cerium concentrations are
labeled next to each data point. (b) Carrier concentration
correction factor ($\frac{\rho_{0}(x)}{\rho_{0}(0.17)}$) versus
change in R$_{H}$ at T = 2.5 K for optimally annealed cerium-doped
samples. (c) $\Delta$T$_{c}$ versus $\Delta\rho_{0}$ for the
irradiated sample. All of the dashed lines are fits to the data.}
\label{AnalysisFig}
\end{figure}

Equation~\ref{deltaresidual} tells us that we are able to
\textit{separate} the effects of oxygen on the residual resistivity
\textit{if} we can eliminate $\frac{\Delta n}{n}$ in the second
term. We calculate $\Delta\rho_{0}(O_{2})$ for each oxygen sample
from the raw data by using the sample annealed at 10$^{-4}$ Torr as
the reference $\rho_{0}$. This sample is chosen as the reference
because we are interested at looking at the changes due to oxygen
within the oxygenated samples and this sample most resembles the
``optimally annealed'' sample in terms of R$_{H}$ and $\rho_{0}$.
The doping term (first term in eq.~\ref{deltaresidual2}) is
determined from previously published data~\cite{Dagan} on optimally
annealed samples, where the Hall coefficient and residual
resistivities for various cerium dopings are known
(Fig.~\ref{AnalysisFig}(a)). From this data we determine the
expected change in the residual resistivity for a given change in
Hall coefficient, $\Delta\rho_{0}(R_{H})$, using the x = 0.17 cerium
doping as our reference. We subtract this from
$\Delta\rho_{0}(O_{2})$ giving a quantity we will call
$\Delta\rho_{0,uncorrected}(disorder)$. This term is not quite the
disorder term in eq.~\ref{deltaresidual2} since we need to eliminate
the carrier concentration dependence. We can determine this
dependence, (1-$\frac{\Delta n}{n}$) in eq.~\ref{deltaresidual}, by
taking the ratios of the residual resistivities of previously
reported \emph{cerium-doped} samples and plotting them as a function
of the change in R$_{H}$ from the x = 0.17 composition
(Fig.~\ref{AnalysisFig}(b)). This can be easily seen if we let the
residual resistivity of the x = 0.17 be $\rho_{0}$(0.17) =
$\frac{m*}{ne^{2}\tau_{0}}$ and all other \emph{cerium dopings}
represented by $\rho_{0}$(x) = $\frac{m*}{(n+\Delta
n)e^{2}\tau_{0}}$. Here we assume $\tau_{0}$ does not depend on
cerium-doping.~\cite{Lin} The (1-$\frac{\Delta n}{n}$) factor can
now be determined from Fig.~\ref{AnalysisFig}(b) for a given
$\Delta$R$_{H}$ within the oxygenated samples. This factor is then
divided out of $\Delta\rho_{0,uncorrected}(disorder)$. We are now
left with the term in equation~\ref{deltaresidual2} due to disorder,
i.e. $\Delta\rho_{0}(disorder)$. This is the crucial term that we
will need in the next step to determine how T$_{c}$ is affected by
disorder in the two more oxygenated films.

\begin{figure}
\centerline{\epsfig{file=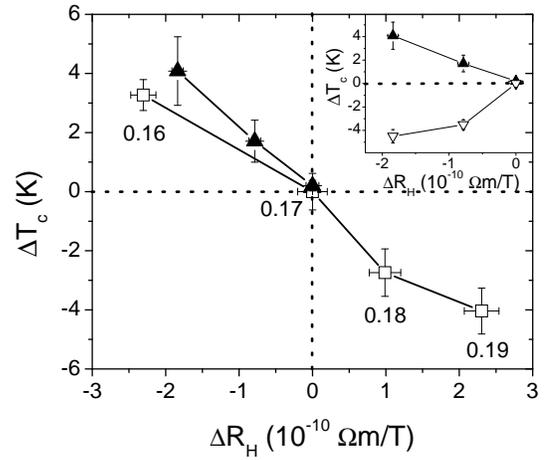,clip=,silent=,width=2.75in}}
\caption{Change in T$_{c}$ versus change in R$_{H}$ at T = 2.5 K.
Doping contributions to the change in T$_{c}$,
$\Delta$T$_{c}$(R$_{H}$), of the x = 0.17 oxygenated samples after
the analysis described in the text ($\blacktriangle$). Optimally
annealed cerium-doped samples ($\square$). Inset shows the raw data
from the oxygenated samples ($\bigtriangledown$) along with
$\Delta$T$_{c}$(R$_{H}$) from the analysis.} \label{FinalFig}
\end{figure}

We use the irradiation data to make a correlation between
$\Delta\rho_{0}$ and the change in T$_{c}$ ($\Delta$T$_{c}$) due to
disorder, in order to determine the expected change in T$_{c}$ of
the oxygenated samples due to disorder. We assume that the change in
T$_{c}$ can be written in the same fashion as
eq.~\ref{deltaresidual2}.
\begin{equation}\label{deltaTc}
  \Delta T_{c}(O_{2}) = \Delta T_{c}(R_{H}) + \Delta T_{c}(disorder)
\end{equation}
The disorder term on the right hand side is determined from the
irradiation data, shown in Fig.~\ref{ResFig}(c) and summarized in
Fig.~\ref{AnalysisFig}(c). With this plot we can now determine
$\Delta$T$_{c}(disorder)$ for each $\Delta\rho_{0}(disorder)$
calculated in the previous paragraph for the two more oxygenated
samples. We subtract $\Delta$T$_{c}(disorder)$ from
$\Delta$T$_{c}(O_{2})$, as determined from Fig.~\ref{ResFig} using
the 10$^{-4}$ Torr sample as the reference, for the two more
oxygenated samples. The result is the contribution to the change in
T$_{c}$ of the oxygenated samples from a change in carrier
concentration, i.e. $\Delta$T$_{c}(R_{H})$ in eq.~\ref{deltaTc}. We
plot our results in Fig.~\ref{FinalFig} along with the data from
cerium-doped samples.

The trend in the oxygenated samples, after our analysis, is
consistent with the trend in the cerium-doped samples, i.e. T$_{c}$
\emph{increases} as R$_{H}$ evolves toward optimal doping (x =
0.15). This is the interesting, and most important, new result of
the research presented in this paper. We see from
Fig.~\ref{FinalFig} that the positive contribution to T$_{c}$ from
hole-doping in the oxygenated samples ($\blacktriangle$ data) is
overshadowed by the negative contribution due to the disorder
introduced by the oxygenation (inset $\bigtriangledown$ data). This
new finding explains why changing the oxygen content in x $\neq$
0.15 samples never results in the maximum T$_{c}$ ($\approx$ 22 K)
of x = 0.15 samples.

We have shown that oxygen has an effect on the properties of PCCO
that can be separated into two parts: disorder and doping. Based on
this result, we now present a possible explanation for the
long-standing puzzle of why oxygen reduction is needed to produce
superconductivity in the \textit{n}-doped cuprates. We will
speculate on the relation between superconductivity and
antiferromagnetism, as well as the lattice sites where oxygen is
removed during reduction.

The overall effect of adding oxygen to a superconducting PCCO (x =
0.17) sample is similar to ion-irradiation, with regards to
disorder. However, irradiation and oxygenation are not expected to
have the same effect on antiferromagnetism. It has been clearly
shown that T$_{N}$ increases as the oxygen content increases (from
an optimal reduction)~\cite{Kang,Mang,Lobo} in the \textit{n}-doped
cuprates. In contrast, one would expect T$_{N}$ to decrease upon
irradiation.~\cite{IrradiationNeel} To that extent, our data support
the conjecture that the suppression of T$_{c}$ by oxygenation is
\textit{primarily disorder-driven} and is not related to any
competing long-range antiferromagnetic order at this cerium-doping.
Conversely, oxygen reduction is necessary to minimize the disorder
which is responsible for inhibiting superconductivity.

This suppression of T$_{c}$ by disorder gives some insight into
where oxygen is removed during the reduction process. Let us look at
this problem from the other perspective and consider the case of
adding oxygen to a reduced sample. In this case, there are three
sites where the oxygen could be entering: the CuO$_{2}$ plane, the
PrO layer, or the apical sites. The first two sites are regular
lattice sites and the reincorporation of oxygen into those sites
would restore the regular lattice potential and reduce disorder. The
last possibility, the apical site, is most likely to increase
disorder as it is predominantly an impurity site in close proximity
to the CuO$_{2}$ plane. Irradiation, on the other hand, introduces
disorder mainly by creating vacancies in the CuO$_{2}$
plane.~\cite{Weaver} Since oxygen is not removed from the material
in this process, it must then be displaced into interstitial sites
of which the apical site is a possibility. The disorder from
irradiation then comes from both the in-plane and the interstitial
sites. Our data suggest that this disorder is quantitatively similar
to adding oxygen, which brings us to the speculation that changing
the occupation of the apical (or interstitial) sites influences the
disorder for a given cerium-doping more than disorder from in-plane
vacancies or from the PrO layer. This interpretation of the effect
of out-of-plane disorder is consistent with Fujita \textit{et
al}.~\cite{Fujita} who reported a strong suppression of T$_{c}$ due
to out-of-plane disorder in the hole-doped cuprates,
La$_{2-x}$Sr$_{x}$CuO$_{4}$ and Bi$_{2}$Sr$_{2}$CuO$_{6+\delta}$.

In summary, we have presented Hall and resistivity data on overdoped
(x = 0.17) PCCO thin films subjected to either oxygenation or
ion-irradiation. The results of the analysis demonstrate that oxygen
has both a doping effect and a disorder effect. Of the two terms,
the disorder effect dominates any change in T$_{c}$ when the oxygen
content is changed. Additionally, while we do not know
\textit{exactly} where oxygen is removed during the reduction
process, we conclude that removal of oxygen from the apical sites is
responsible for the reduction of the disorder that inhibits the
appearance of superconductivity in the \textit{n}-doped cuprates.

The authors would like to acknowledge W. Yu, A. Zimmers, A. J.
Millis, and J. W. Lynn for their helpful discussions. This work was
supported by NSF Grant DMR-0352735 and in part by ONR.


\begin{references}
\bibitem{Takagi} H. Takagi, S. Uchida, and Y. Tokura, Phys. Rev. Lett. \textbf{62}, 1197 (1989)
\bibitem{Jiang} W. Jiang, S. N. Mao, X. X. Xi, X. Jiang, J. L. Peng,
T. Venkatesan, C. J. Lobb, and R. L. Greene, Phys. Rev. Lett.
\textbf{73}, 1291 (1994)
\bibitem{Kang} H. J. Kang, P. Dai, H. A. Mook, D. N. Argyriou, V.
Sikolenko, J. W. Lynn, Y. Kurita, S. Komiya, and Y. Ando, Phys. Rev.
B \textbf{71}, 214512 (2005)
\bibitem{Xu} X. Q. Xu, S. N. Mao, W. Jiang, J. L. Peng, and R. L.
Greene, Phys Rev. B \textbf{53}, 871 (1996)
\bibitem{Richard} P. Richard, G. Riou, I. Hetel, S. Jandl, M.
Poirier, and P. Fournier, Phys. Rev. B \textbf{70}, 064513 (2004)
\bibitem{Riou} G. Riou, P. Richard, S. Jandl, M. Poirier, P.
Fournier, V. Nekvasil, S. N. Barilo, and L. A. Kurnevich, Phys. Rev.
B \textbf{69}, 024511 (2004)
\bibitem{Jiang2} W. Jiang, J. L. Peng, Z. Y. Li, and R. L. Greene, Phys. Rev. B
\textbf{47}, 8151 (1993)
\bibitem{Radaelli} P. G. Radaelli, J. D. Jorgensen, A. J. Schultz, J.
L. Peng, and R. L. Greene, Phys. Rev. B \textbf{49}, 15322 (1994)
\bibitem{Woods} S. I. Woods, A. S. Katz, S. I. Applebaum, M. C. de
Andrade, M. B. Maple, and R. C. Dynes, Phys. Rev. B \textbf{66},
014538 (2002)
\bibitem{Woods2} S. I. Woods, A. S. Katz, M. C. de Andrade, J.
Herrmann, M. B. Maple, and R. C. Dynes, Phys. Rev B \textbf{58},
8800 (1998)
\bibitem{Weaver} B. D. Weaver, G. P. Summers, R. L. Greene, E. M.
Jackson, S. N. Mao, and W. Jiang, Physica C \textbf{261}, 229 (1996)
\bibitem{Tolpygo} S. K. Tolpygo, J.-Y. Lin, M. Gurvitch, S. Y. Hou,
and J. M. Phillips, Phys. Rev. B \textbf{53}, 12454 (1996)
\bibitem{Fournier} P. Fournier, X. Jiang, W. Jiang, S. N. Mao, T.
Venkatesan, C. J. Lobb, and R. L. Greene, Phys. Rev. B \textbf{56},
14149 (1997)
\bibitem{Gollnik} F. Gollnik and M. Naito, Phys. Rev. B \textbf{58},
11734 (1998)
\bibitem{Padilla} W. J. Padilla, Y. S. Lee, M. Dumm, G. Blumberg, S.
Ono, Kouji Segawa, Seiki Komiya, Yoichi Ando, and D. N. Basov, Phys.
Rev. B \textbf{72} 060511 (2005)
\bibitem{Zimmers} A. Zimmers (private communication)
\bibitem{Dagan} Y. Dagan, M. M. Qazilbash, C. P. Hill, V. N.
Kulkarni, and R. L. Greene, Phys. Rev. Lett. \textbf{92}, 167001
(2004)
\bibitem{Lin} J. Lin and A. J. Millis, Phys. Rev. B \textbf{72}, 214506 (2005)
\bibitem{Mang} P. K. Mang, O. P. Vajk, A. Arvanitaki, J. W. Lynn, and M.
Greven, Phys. Rev. Lett. \textbf{93}, 027002 (2004)
\bibitem{Lobo} R. P. S. M. Lobo, N. Bontemps, A. Zimmers, Y. Dagan,
R. L. Greene, P. Fournier, C. C. Homes, and A. J. Millis, SPIE
Optics and Photonics conference proceedings, 31 July - 4 August
2005, to be published
\bibitem{IrradiationNeel} A lack of literature in regards to experimental
results on the effects of T$_{N}$ upon irradiation leaves us with no
definitive answer as to which direction T$_{N}$ would actually move,
if at all. The conjecture is that T$_{N}$ would reduce because of a
modification of the Cu-Cu exchange interaction energy due to the
disruption of the oxygen and copper sublattices created by vacancies
in the CuO$_{2}$ plane.
\bibitem{Fujita} K. Fujita, T. Noda, K. M. Kojima,H. Eisaki, and S. Uchida,
Phys. Rev. Lett. \textbf{95}, 097006 (2005)
\end{references}
\end{document}